\documentstyle[preprint,prl,aps]{revtex}
\begin{document}
\draft
\title{Path Integral Approach to the Nonextensive Canonical Density Matrix}
\author{E. K. Lenzi$^{1}$, L. C. Malacarne$^{2}$ and R. S. Mendes$^{2}$
\thanks{ e-mail address:  rsmendes@dfi.uem.br, fax number: 55 44 2634242}}
\address{
$^1$Centro Brasileiro de Pesquisas F\'\i sicas,
R. Dr. Xavier Sigaud 150, \\
22290-180 Rio de Janeiro-RJ, Brazil\\
$^2$Departamento de F\'\i sica, Universidade Estadual de Maring\'a, \\
Av. Colombo 5790, 87020-900 Maring\'a-PR, Brazil \\
}
\date{\today }
\maketitle
\begin{abstract}
Feynman's path integral is herein generalized  to the nonextensive
canonical density matrix  based on Tsallis entropy. This
generalization is done in two ways by using unnormalized and
normalized constraints. Firstly, we consider the path integral
formulation with unnormalized constraints, and this generalization
is worked out through two different ways, which are shown to be
equivalent. These formulations with unnormalized constraints are
solutions to two generalized Bloch equations proposed in this
work. The first form of the generalized Bloch equation is linear,
but with a temperature-dependent effective Hamiltonian; the second
form is nonlinear and resembles the anomalous correlated diffusion
equation (porous medium equation). Furthermore, we can extend
these results to the prescription of field theory using integral
representations. The second development is dedicated to analyzing
the path integral formulation with normalized constraints.
 To illustrate the methods introduced here,
 we analyze the free particle case and a non-interacting scalar
field.  The results herein obtained are expected to be useful in
the discussion of generic nonextensive contexts.
\end{abstract}
\date{\today}
\section{INTRODUCTION}
Path integral techniques constitute a powerful tool, employed to
analyze countless physical situations, for instance, in standard
statistical mechanics (Boltzmann-Gibbs).
On the other hand, there is nonextensive
behavior common in many branches of physics, for example, systems
with long-range (gravitational) interactions, long-time memory,
and fractally structured space-time. This nonextensive behavior
indicates that the standard statistical mechanics and
thermodynamics need some enlargement. Consequently, the path
integral employed in standard statistical mechanics also needs
some generalizations in order to accommodate the nonextensive
effects.

Recently, a theoretical tool based on nonextensive entropy
(Tsallis entropy)\cite{T1} has successfully been applied in many
situations, for example, L\'{e}vy-type anomalous
superdiffusion\cite{6}, Euler turbulence \cite{7},
self-gravitating systems\cite{8}, cosmic background radiation
\cite{9}, peculiar velocities of galaxies\cite{10a}, anomalous
relaxation through electron-phonon interaction\cite {10c},
ferrofluid-like systems\cite{11}, and nonlinear dissipative
dynamical systems\cite{11a}. In this context, it is very important
to understand  the main properties of the generalized statistical
mechanics based on Tsallis entropy more deeply. In particular, it
is interesting to develop basic tools for calculations such as the
semi-classical approximation\cite{ev}, perturbation \cite{le,le1}
and variational methods\cite{le,le1,plast}, linear response theory
\cite{10b}, Green functions\cite{RAJA,abe3}, non-Gaussian
integrals \cite{R2}, and path integral, which is the subject of
this work.

We develop two path integral formalisms and Bloch equations based
on Tsallis statistics with unnormalized constraints in Section II.
The first form of the generalized Bloch equation is linear, but
with an effective Hamiltonian, which has a temperature dependence,
and the other equation resembles the anomalous diffusion equation
(porous medium equation). The  {\it path integral} solutions to
these two different but equivalent {\it generalized Bloch
equations} are also obtained in Sections II as well as the
extension to the field theory. The path integral is first
formulated for effective Hamiltonian case. In the other case, we
use integral representations to formulate the path integral. To
illustrate theses methods, the free particle case for $q>1$ and
$q<1$ and the free scalar field bosonic  are studied. Section III
is dedicated to analyzing the {\it path integral} formulation
obtained with normalized constraints and its extension to quantum
field theory, where an example is exhibited. Conclusions are
presented in Section IV.


\section{PATH INTEGRAL BASED IN UNNORMALIZED CONSTRAINTS}

Let us start this section with some remarks about Bloch equation
and nonextensive Tsallis statistics in order to introduce the
generalized  Bloch equations. After this presentation, we obtain
the path integral formulation for the solutions of these Bloch
equations, giving as example the free   particle case.
Furthermore, we apply these developments for a brief discussion
about quantum scalar fields.


{\subsection  {BLOCH EQUATION}}

The usual density matrix
\begin{eqnarray}
\hat{p}_{1}(\beta)= {\mbox e}^{- \beta \hat{H} }/Z_1 \;\;\;\;\; \left({\mbox
{Tr}}\;\hat{p}_{1}(\beta)=1\right)  \label{b1}
\end{eqnarray}
($Z_1={\mbox {Tr}}\;e^{- \beta \hat{H}} $ is the usual partition function)
can be obtained directly from a maximum entropy principle\cite{Ba}, and the
{\it unnormalized} matrix
\begin{eqnarray}
\hat{\rho}_1(\beta)= \exp (- \beta \hat{H} )  \label{b102}
\end{eqnarray}
obeys the Bloch equation\cite{Bl} (see also Ref. \cite{feynman1} p. 48)
\begin{eqnarray}
-\frac{\partial \hat{\rho_1}}{\partial \beta} = \hat{H}
\hat{\rho_1} \; \label{b2}
\end{eqnarray}
with the initial condition $\hat{\rho}_1(\beta=0) = {\bf 1 } $.
When $\hat{\rho}_1(\beta)$ is calculated and subsequently $Z_1$,
we can obtain the usual free energy $F_1=-(1/\beta) \ln Z_1$.

Recently, the density matrix (\ref{b1}) was generalized in the
nonextensive context\cite{T1,CT,plass}, {\it i.e.},
\begin{eqnarray}
\hat{p}_q (\beta) = [ 1-(1-q)\beta \hat{H}]^{1/(1-q)}/Z_q \; ,
\label{b3}
\end{eqnarray}
where $Z_q={\mbox {Tr}} [ 1-(1-q)\beta \hat{H}]^{1/(1-q)}$ is the
generalized partition function. In Eq.(\ref{b3}) we are supposing
that $ 1-(1-q)\beta E_n \geq 0$, where $E_n$ are the eingenvalues
of $\hat{H}$. When this condition is not satisfied, we employ
$p(E_n)=0$, {\it i.e.}, a cut-off is used so that $p_q(E_n)$
remains positive. By using $Z_q$, we can obtain the generalized
free energy $F_q=-(1/\beta) (Z_q^{1-q} -1)/(1-q)$. These
generalizations were obtained by employing the maximum entropy
principle with the Tsallis entropy,
\begin{eqnarray}
S_q = - k \; {\mbox {Tr}}\; \frac{\hat{\rho} -\hat{\rho}^{q}}{1-q} \; ,
\label{b4}
\end{eqnarray}
and the constraints $\langle \hat{H} \rangle_q = {\mbox {Tr}}
\;\hat{\rho}_q^q \hat{H}$ and ${\mbox {Tr}} \;\hat{\rho}_q = 1$
($k$ is a positive constant). The parameter $q\in {\bf R}$ gives
the degree of nonextensivity, and in the limit $q \rightarrow 1$
the usual statistical mechanics is recovered. Note that the
constraint of the energy is unnormalized, since $\langle 1
\rangle_q= {\mbox {Tr}}\hat{\rho}_q ^q \neq 1$. We return to this
point at Section IV (see  Ref. \cite{NORMA} for a complete
discussion about the choice of the  constraints).

Following the previous discussion about the Block equation, we
employ the unnormalized matrix
\begin{eqnarray}
\hat{\rho}_q (\beta) = [ 1-(1-q)\beta \hat{H}]^{1/(1-q)} \; .
\label{b5}
\end{eqnarray}
This matrix satisfies alternatively  the equations
\begin{eqnarray}
-\frac{\partial \hat{\rho}_q}{\partial \beta}
= \frac{\hat{H}}{ 1-(1-q)\beta
\hat{H} } \; \hat{\rho}_q \;\;,  \label{b61}
\end{eqnarray}
\begin{eqnarray}
-\frac{\partial \hat{\rho}_q}{\partial \beta} = \hat{H}
\hat{\rho}_q^q \; ,\label{b6}
\end{eqnarray}
with the initial condition $\hat{\rho}_q(\beta=0) = {\bf 1}$.
These equations generalize the Bloch equation in the nonextensive
Tsallis context. In fact, the generalized Bloch equation
(\ref{b61}) is linear in $\hat{\rho} _q$ and has the same form of
the Eq. (\ref{b2}), if we employ the effective Hamiltonian
$\hat{H}_{eff}=\hat{H}[1-(1-q)\beta \hat{H} ]^{-1}$. On the other
hand,  the generalized Bloch equation (\ref{b6}) has some formal
resemblace to the anomalous correlated diffusion equation
\cite{pplas}. In the following, we develop the path integral
solution for both generalized Bloch equations.

{\subsection  {PATH INTEGRAL FORMULATION WITH $H_{eff}$}}

Since Eqs.(\ref{b2}) and (\ref{b61}) have the same form, the path
integral representation for $\hat{\rho}_q$ has the same structure
of  the usual path integral. Thus, the path integral for
$\hat{\rho}_q$ can be written as
\begin{eqnarray}
& &  \rho _q (x,x^{\prime}; \beta)  =\langle
x|\hat{\rho}_q(\beta)|x^{\prime}\rangle  \nonumber
\\ &=& \lim_{N \rightarrow \infty} \int \; .\; .\; .\int \left[
\prod_{n=1}^{N-1} {\mbox d}x_n \right]
 \left[ \prod_{n=1}^{N} \frac{{\mbox d}%
p_n}{2\pi\hbar} \right] \exp
\left\{ \frac{1}{\hbar} \sum_{n=1}^{N} [i\; p_n
(x_n - x_{n-1}) - \epsilon H_{eff}(p_n, \bar{x}_n, n\epsilon) ] \right\}
\nonumber \\
& = & \int {\cal D}x {\cal D}p \; \exp \left[\frac{1}{\hbar} \int_0^{\beta}
\left(i\; p \dot{x} - H_{eff} \right) {\mbox d}\tau \right ] \; ,
\label{3.1}
\end{eqnarray}
 where $\bar{x}_n = (x_{n+1}-x_n)/2$,  $x=x_0$, $x'=x_N$ and
$\epsilon= \beta/N$. In this path integral representation, we are
considering that the Hamiltonian $\hat{H}$ as well as
$\hat{H}_{eff}$ are Weyl ordered (if this is not the case, the
discussion has to be redone along the present lines), and
consequently the midpoint prescription was employed\cite{schu}. In
general, this path integral is not easy to calculate because the
Hamiltonian $\hat{H}_{eff}$($=$ $\hat{H}[1-(1-q)\beta \hat{H}
]^{-1}$) has an unusual nonlinear dependence on the moments and
coordinates. However, as we shall see, it is possible to calculate
$\hat{\rho}_q$ for the free particle case.  Simple textbook
examples, such as free particle, the harmonic oscillator, the
non-interacting scalar field, etc., do not exibit any kind of
nonextensive behavior. It is not necessary to introduce a
generalized statistical mechanics to deal with these systems.
However, these simple examples are very useful in order to
illustrate how the formalism works. And  they are even more
instructive when they provide exactly solvable cases. In this
case,  considering the free particle case, we first perform the
integration in the coordinates,
\begin{eqnarray}
& & \rho_q (x,x^{\prime}; \beta) = \int \frac{{\mbox
d}p}{2\pi\hbar} \exp\left[\frac{ip}{\hbar}(x-x^{\prime})\right]
\lim_{N \rightarrow \infty} \int \; .\; .\; .\int \left[
\prod_{n=1}^{N-1} {\mbox d}p_n \delta (p_n -p_{n-1}) \right]
\times  \nonumber \\ &\times &\exp \left\{ - \epsilon
\sum_{n=1}^{N} \left[ \frac{ p_n^2/2m }{ 1-\epsilon n (1-q)
p_n^2/2m} \right] \right\} \; .  \label{3.2}
\end{eqnarray}
By using the identity (see appendix A)
\begin{eqnarray}
\ln(1+x)=\lim_{N\rightarrow \infty}\sum_{n=1}^{N}
\left(\frac{ x}{N }\right)
\frac{1}{1 + \frac{nx }{N}}\; ,
\end{eqnarray}
and considering the case $q>1$ with $\beta >0$, we verified that
Eq. (\ref{3.2}) reduces to
\begin{eqnarray}
\rho_q(x,x^{\prime};\beta) &=& \int_{-\infty}^{\infty}
\frac{{\mbox d}p}{%
2\pi \hbar} \exp\left[\frac{ip}{\hbar}(x-x^{\prime})\right]
\left[1-(1-q)\beta \frac{p^2}{2m}\right]^{1/(1-q) }  \nonumber \\
& =& \frac{1}{\Gamma(\frac{1}{q-1})}
\left[ \frac{2 m }{\pi (q-1)\beta \hbar
^2} \right]^{1/2}
\left[ \left( \frac{m(x-x^{\prime})^2}{2 (q-1) \beta \hbar
^2} \right)^{1/2} \right]^{1/(q-1) -1/2 } \times  \nonumber \\
& \times & K_{1/(q-1)-1/2}
\left[ \left ( \frac{2m(x-x^{\prime})^2}{(q-1)
\beta \hbar^2 } \right )^{1/2} \right] \; .  \label{d2}
\end{eqnarray}
In the last step, we used a known integral (for instance, see Ref.
\cite{gr1} p. 321) and relation between the Wittaker function
$W_{0,\; \nu}$ and the modified Bessel function of second kind
$K_{\nu}$.

In the case $q<1$ with $\beta >0$, that has a cut-off, the integration
limits of the moments are limited and the
path integral (\ref{3.2}) reduces
to the first part of Eq. (\ref{d2}) with $p$ satisfying the condition $
1-(1-q)\beta {p^2}/{2m} \geq 0$. Thus, when the remaining
integration is performed (see Ref. \cite{gr1} p. 321), we obtain
\begin{eqnarray}
\rho_q(x,x^{\prime};\beta) &=& \int_{-\left[ \frac{2m}{(1-q) \beta}
\right]^{1/2}}^{\left[\frac{2m}{(1-q) \beta}\right]^{1/2}}
 \frac{{\mbox d}p}{
2\pi \hbar} \exp\left[\frac{ip}{\hbar}(x-x^{\prime})\right]
\left[1-(1-q)\beta \frac{p^2}{2m}\right]^{1/(1-q) }  \nonumber \\
& = & \Gamma\left( \frac{1}{1-q} +1\right)
\left[ \frac{m}{2 \pi (1-q)\beta
\hbar ^2}\right]^{1/2} \left[\left(\frac{2(1-q)\beta \hbar ^{2}}{m
(x-x^{\prime})^2} \right)^{1/2} \right]^{1/(1-q)+1/2 } \times
 \nonumber \\
&\times &J_{1/(1-q)+1/2} \left[ \left( \frac{2m(x-x^{\prime})^2}{
(1-q)\beta\hbar^2} \right)^{1/2} \right] \; .
\end{eqnarray}

The partition function can be obtained for $q>1$ and $q<1$ by taking the
trace of $\rho_q(x,x^{\prime}; \beta)$. In fact, for $q>1$ we obtain
\begin{eqnarray}
Z_q &=& \int_{-L/2}^{L/2} dx \;\rho_q(x,x; \beta)  \nonumber \\
&=& L\left[ \frac{m}{2 \pi (q-1) \beta \hbar^2} \right]^{1/2}
\frac{\Gamma(
\frac{1}{q-1} -\frac{1}{2})} {\Gamma(\frac{1}{q-1})} \; ,
 \label{d.4}
\end{eqnarray}
and for $q<1$ we have
\begin{eqnarray}
Z_q= L \left[ \frac{m}{2 \pi (1-q) \beta \hbar^2} \right]^{1/2}
\frac{\Gamma(
\frac{1}{1-q} +1) }{\Gamma(\frac{1}{1-q} + \frac{3}{2})} \; ,
 \label{d.5}
\end{eqnarray}
where $L$ is the length of the integration region (box size).
Because $L$ is
large the expressions (\ref{d.4}) and (\ref{d.5}) reduce to the
generalized
classical partition function.

${ }$

{\subsection  {PATH INTEGRAL FORMULATION WITH INTEGRAL REPRESENTATIONS}}

${ }$
Let us consider the Hilhorst's identity (private communication to Tsallis
\cite{Hi})
\begin{eqnarray}
[ 1-(1-q)\beta \hat{H}]^{1/(1-q)}= \frac{1}{ \Gamma \left(\frac{1}{q-1}
\right)} \int_0^{\infty} {\mbox d}v \;v^{\frac{1}{q-1}-1} {\mbox e}^{-v}
\exp(- v (q-1)\beta \hat{H}) \; .  \label{b7}
\end{eqnarray}
This operator identity is essentially a direct application of the
usual integral representation of the gamma function. Indeed, if
both sides of the Eq. (\ref {b7}) are applied on an eigenstate of
$\hat{H}$ with $q>1$, $\beta >0$, and $ E_n >0 $ ($E_n$ is an
arbitrary eigenvalue of $\hat{H}$) the identity (\ref{b7}) reduces
to the usual integral representation of the gamma function. From
Eq. (\ref{b7}) the path integral representation for
$\hat{\rho}_q(\beta) $ can be obtained. In fact, it is necessary
to employ only the path integral representation for
$\hat{\rho}_1(v (q-1) \beta)$ in the last part of Eq. (\ref {b7}).
Thus,
\begin{eqnarray}
\rho_q(x,x^{\prime}; \beta) = \frac{1}{ \Gamma \left(\frac{1}{q-1}\right)}
\int_0^{\infty}{\mbox d}v \; v^{\frac{1}{q-1}-1} {\mbox e}^{-v}
\rho_1(x,x^{\prime}; v (q-1)\beta ) \; ,  \label{b8}
\end{eqnarray}
where $\rho_1(x,x^{\prime}; v (q-1)\beta )$ must be replaced by
its usual (or holomorphic) path integral representation.
Furthermore, if we take the trace of Eq. (\ref{b8}), we obtain the
connection between the generalized and usual partition
functions\cite{Hi}.

As illustration , let us apply Eq.(\ref{b8}) to obtain the
generalized density matrix for a free particle in the
one-dimensional case. For this example, the usual density matrix
in the coordinate representation is (for instance, see Ref.
\cite{feynman1} p. 49)
\begin{eqnarray}
\rho_1(x,x^{\prime};\beta) = \sqrt{\frac{m}{2\pi \hbar^2 \beta} } \exp
\left[ - \left( \frac{m}{2 \hbar^2 \beta}\right) (x-x^{\prime})^2 \right ]
\; .  \label{b81}
\end{eqnarray}
When this expression is replaced in Eq. (\ref{b8}) and the
integral is calculated (see Ref. \cite{gr1} p. 935), we obtain, as
expected,  Eq. (\ref{d2}).

The integral transformation (\ref{b7}) can not be applied for
$q<1$, because the integral is not convergent. However, for $q<1$
the Prato's identity
\begin{eqnarray}
\lbrack 1-(1-q)\beta \hat{H}]^{1/(1-q)}=\frac{i}{2\pi }\Gamma \left( \frac{
2-q}{1-q}\right) \int_{C}{\mbox d}v\;(-v)^{-\frac{2-q}{1-q}}{\mbox e}
^{-v}\exp (v(1-q)\beta \hat{H})\;
\label{p1}
\end{eqnarray}
can be employed\cite{prato}, except for some particular values of
$q$, $(2-q)/(1-q)= {\mbox integer}$. As in the case $q>1$, this
integral representation is a direct consequence of  another gamma
function representation (for instance, see Ref. \cite{gr1} p.
935). The contour integration can be chosen in order to eliminate
the condition $(2-q)/(1-q)\neq {\mbox integer}$, see for instance,
Refs. \cite{le1} and \cite{negro}. Thus, we can formulate a path
integral for $q<1$ using Eq.(\ref{p1}). In fact, it is sufficient
to replace $\exp (v(1-q)\beta \hat{H})$ by its path integral
representation in Eq. (\ref{p1}) to obtain the desirable path
integral representation for $\rho _{q}(x,x^{\prime }; \beta )$
with $q<1$.

${ }$

{\subsection   {EXTENSION TO THE FIELD THEORY FORMALISM }}

${ }$
We can use the previous  path integral prescription to study
some aspects of the Tsallis' nonextensive thermostatistics as
applied to field theory. Let us start our discussion about
nonextesive statistical field theory by considering the scalar
field. In this example, we obtain the generalized partition
function and the pressure attributed to the ground state of  a
free scalar field. The Lagrangian density is
\begin{eqnarray}
{\cal L}= \frac{1}{2} \partial_{\mu}\phi\partial^{\mu}\phi -
\frac{1}{2} m^2 \phi^2\; ,
\end{eqnarray}
with $c=1$, $\hbar=1$, and consequently the Hamiltonian
density is
\begin{eqnarray}
{\cal H}= \frac{1}{2} \pi^2 + \frac{1}{2}(\nabla \phi)^2+
\frac{1}{2} m^2 \phi^2 \;.
\label{H}
\end{eqnarray}
By taking the trace of the Eq.(\ref{p1}), the generalized
partition function for the free field, we get
\begin{eqnarray}
Z_q=  \int_{C}{\mbox d}v\;K_q^{(1)}(v) \int   \cdots \int {\cal
D}\phi {\cal D} \pi \; e^{\int_{0}^{\beta^{*}} {\mbox {d}} \tau
\int {\mbox {d}}^3 x \left(\phi\cdot \pi - {\cal H}(\phi,
\pi)\right) }\; ,
\end{eqnarray}
with $\beta^{*}=(1-q)(-v)\beta$ and  $K_q^{(1)} = \frac{i}{2 \pi}
\Gamma \left( \frac{2-q}{1-q}\right)
(-v)^{-\left(\frac{2-q}{1-q}\right) } e^{-v}$. This partition
function, after the momentum integration  reduces to
\begin{eqnarray}
Z_q=  \int_{C}{\mbox d}v\;K_q^{(1)}(v) \int   \cdots
\int {\cal D}\phi \;
e^{\int_{0}^{\beta^{*}} {\mbox {d}}
\tau \int {\mbox {d}}^3 x {\cal L} (\phi,
\partial_{\mu}{\phi})}\;.
\end{eqnarray}
By substituting the Lagrangian density and following the usual
 calculation
\cite{kapusa}, we obtain for the generalized
partition function the following expression
\begin{eqnarray}
Z_q=\int_{C}{\mbox d}v\;K_q^{(1)}(v) \exp \left(
-\frac{1}{2}(\beta^*)^2\sum_{n}\sum_{p}(\omega_n^2 + \omega^2)
\right)\; , \label{ZF}
\end{eqnarray}
where $\omega=(\vec{p}^{\;2}+m^2)^{1/2}$ and $\omega_n=2 \pi
n/\beta^*$. Performing the sum over $n$, the above expression can
be rewritten as
\begin{eqnarray}
Z_q\!=\! \int_{C}{\mbox d}v\;K_q^{(1)}(v)\! \exp \left \{\!-V\!
\int \frac{{\mbox {d}^3} p}{(2 \pi)^3}\! \left[\frac{1}{2}\beta
(1-q)(-v)\omega +\ln\left( 1-e^{ -\beta(1-q)(-v)\omega} \right)
\right]      \right \} \;. \label{z1q}
\end{eqnarray}
By taking $m\rightarrow 0$ this expression recovers essentially
the generalized partition function for the blackbody
radiation\cite{negro}. In the following, as illustration, we
calculate the generalized pressure attributed to the ground state.
In this case, we use the definition of the generalized pressure
\cite{negro,30a}
\begin{eqnarray}
P_q =\frac{1}{\beta} \frac{\partial}{\partial V}
\left(\frac{Z_q^{1-q} -1}{q-1} \right)\;. \label{pressure}
\end{eqnarray}
 In this way, by using Eqs.
(\ref{z1q}) and (\ref{pressure}) with $T=1/\beta$ sufficiently
small (a detail discussion of this limit is presented in Ref.
\cite{le1}), we obtain
\begin{eqnarray}
P_q &=& \frac{1}{Z_q^q} \! \int_{C}{\mbox d}v\;K_q^{(2)}(v) \!
\int \frac{{\mbox {d}}^3 p} {(2 \pi)^3}\! \left[\frac{1}{2}\omega
+\frac{1}{\beta} \ln\left(1-e^{ \beta(1-q)(-v) \omega}
\right)\right] \times \nonumber \\ &\times& \exp \left \{-V\! \int
\frac{{\mbox {d}}^3 p}{(2 \pi)^3}\! \left[ \frac{1}{2}\beta
(1-q)(-v)\omega+ \ln \left( 1- e^{-\beta(1-q)(-v)\omega} \right)
\right] \right \} \nonumber \\ &=& \int \frac{{\mbox {d}}^3 p}{(2
\pi)^3} \omega  \;, \label{pressao}
\end{eqnarray}
where
\begin{equation}\label{k2}
K_q^{(2)}(v)= \frac{\Gamma\left(\frac{1}{1-q}\right)}{2 \pi i}
e^{-v}(-v)^{-1/(1-q)}\; .
\end{equation}

\noindent
 At this point, we remark that the result for the
generalized pressure of the ground state coincides with the same
$q=1$
 case \cite{kapusa}. In general, the ground state properties can
be obtained by using the small temperature limit of the
generalized statistics independently of the $q$ value.

${ }$

\section{Path integral based on normalized constraints}

${ }$ As in the above discussion, we can formulate the path
integral for nonextensive systems using another version of the
generalized statistics with different constraints(a detailed
discussion of the choice of the constraints is presented in Ref.
\cite{NORMA}). Considering the case with the normalized constraint
$ \langle \langle \hat{H} \rangle \rangle_{q} = {\mbox {Tr}}
\hat{\tilde{\rho}}_{q}^q  \hat{H} /{\mbox {Tr}}
\hat{\tilde{\rho}}_q^q $, instead  of $ \langle \hat{H}
\rangle_{q} = {\mbox {Tr}} \hat{\tilde{\rho}}_{q}^q  \hat{H} $, we
obtain that
\begin{equation}\label{N1}
  \hat{\tilde{p}}_q =\frac{\hat{\tilde{\rho}}_q}{\tilde{Z}_{q}}\; ,
\end{equation}

\noindent
 with $\hat{\tilde{\rho}}_q=\left(1-(1-q)
\tilde{\beta}(\hat{H}- \langle \langle \hat{H} \rangle \rangle_{q}
)\right)^{1/(1-q)}$, $\tilde{Z}_{q} = {\mbox
{Tr}}\hat{\tilde{\rho}}_q$, and $\tilde{\beta}= \beta/{\mbox
{Tr}}\hat{\tilde{p}}^{\; q}_q$. In the same way of the
unnormalized constraint case,  we can express  Eq.(\ref{N1})
through the path integral,
\begin{eqnarray}
\tilde{\rho}_q (\phi ,\phi^\prime , \beta)= \int_C {\mbox {d}}u
\tilde{K}_q^{(1)}(v) \int \cdots \int {\cal D}\phi {\cal D}\pi
\exp \left[ \int_0^{\overline{\beta}} \int {\mbox d}\tau {\mbox
d}^3 x (\phi \cdot \pi -{\cal H}(\phi, \pi))\right ] \;
,\label{N2}
\end{eqnarray}
where $\overline{\beta}=(1-q) \tilde{\beta}(-v)$ and
$\tilde{K}_q^{(1)}(v)= K_q^{(1)}(v) e^{-v(1-q)\tilde{\beta}
\langle \langle \hat{H} \rangle \rangle_{q}}$.

To exemplify the normalized formalism, let us   consider the free
particle case,
\begin{eqnarray}
\tilde{\rho}_q (x,x',\beta )= \int_C {\mbox {d}}v\;
\tilde{K}_q^{(1)}(v)
\int \cdots \int {\cal D}x {\cal D}p \exp
\left[ \int_0^{\overline{\beta}} {\mbox d}
\tau (x p - H(x,p))\right ] \;.
\end{eqnarray}
\noindent
 Here, the
calculation resembles  the unnormalized case. Thus, we restrict
ourselves  to presenting the main results. For instance, in the
$q<1$ case, after some calculations, the density matrix for one
particle is given by
\begin{eqnarray}
\tilde{\rho}_{q} (x, x', \beta ) &=& \left[ \frac{m{\mbox
{Tr}}\hat{\tilde{p}}^q}{2 \pi (1-q)\beta \hbar ^2}
\right]^{1/2}
 \left[\left(\frac{2(1-q)\beta \hbar ^{2}}{{m
{\mbox {Tr}}\hat{\tilde{p}}^q} (x-x^{\prime})^2} \left(1+
\frac{(1-q)\beta}{{\mbox {Tr}}\hat{\tilde{p}}^q} \langle \langle
\hat{H} \rangle \rangle_{q} \right) \right)^{\frac{1}{2}}
\right]^{\frac{1}{1-q}+\frac{1}{2}} \nonumber \\
 &\times&  \Gamma\left( \frac{2-q}{1-q}\right)J_{1/(1-q)+1/2}
\left[ \left( \frac{2m(x-x^{\prime})^2{{\mbox
{Tr}}\hat{\tilde{p}}^q}}{ (1-q)\beta\hbar^2}\left(1+
\frac{(1-q)\beta}{{\mbox {Tr}}\hat{\tilde{p}}^q} \langle \langle
\hat{H} \rangle \rangle_{q} \right)\right)^{1/2} \right]
, \label{N5}
\end{eqnarray}
and, for the $q>1$, it reduces to
\begin{eqnarray}
\tilde{\rho}_q (x, x',\beta )&=& \left[ \frac{2 m {\mbox
{Tr}}\hat{\tilde{p}}^q}{\pi (q-1)\beta \hbar ^2} \left(1+
\frac{(1-q)\beta}{{\mbox {Tr}}\hat{\tilde{p}}^q} \langle \langle
\hat{H} \rangle \rangle_{q} \right) ^{\frac{3-q}{1-q}}
\right]^{\frac{1}{2}}
\left[ \left( \frac{m(x-x^{\prime})^2{\mbox
{Tr}}\hat{\tilde{p}}^q}{2 (q-1) \beta \hbar ^2} \right)^{1/2}
\right]^{\frac{1}{q-1} -\frac{1}{2} } \nonumber \\ & \times &
\frac{1}{\Gamma(\frac{1}{q-1})}
K_{1/(q-1)-1/2} \left[ \left ( \frac{2m(x-x^{\prime})^2{\mbox
{Tr}}\hat{\tilde{p}}^q}{(q-1) \beta \hbar^2 } \left(1+
\frac{(1-q)\beta}{{\mbox {Tr}}\hat{\tilde{p}}^q} \langle \langle
\hat{H} \rangle \rangle_{q} \right) \right)^{\frac{1}{2}}
\right]\; .
\end{eqnarray}
Here,  $\tilde{Z}_{q}$ and $\langle \langle \hat{H} \rangle
\rangle_{q}$ for $q<1$ are
\begin{eqnarray}
\tilde{Z}_q&=& \left[ L \left( \frac{2m \pi}{(1-q) \beta h^2}
\right)^{1/2} \frac{\Gamma \left( \frac{2-q}{1-q} \right)}
 {\Gamma
\left( \frac{2-q}{1-q} + \frac{1}{2} \right)}
(1+(1-q)\frac{1}{2})^{1/(1-q)+1/2} \right]^{2/(1+q)}
 \nonumber \\
\langle \langle \hat{H} \rangle \rangle_q
&=& \frac{1}{2 \beta}
{\mbox {Tr}}\hat{\tilde{p}}_q^q \; ,
\end{eqnarray}
and, for the case $q>1$, are
\begin{eqnarray}
\tilde{Z_q}&=& \left[L \left( \frac{2m \pi}{(q-1) \beta h^2 }
\right)^{1/2} \frac{\Gamma \left( \frac{1}{q-1}-\frac{1}{2}
\right)} {\Gamma \left( \frac{1}{q-1} \right)}
(1+(1-q)\frac{1}{2})^{1/(1-q)+1/2} \right]^{2/(1+q)} \nonumber \\
\langle \langle \hat{H} \rangle \rangle_q &=& \frac{1}{2
\beta}{\mbox {Tr}}\hat{\tilde{p}}_q^q\; .
\end{eqnarray}
 In these calculations   the general relation
${\mbox {Tr}} \hat{\tilde{p}}^q_q = \tilde{Z}_{q}^{1-q}$ was
employed.

A useful guide to calculate thermodynamics quantities
 in the normalized  formulation is based on the
 relation\cite{NORMA}

\begin{eqnarray}
\beta^{unnor}= \frac{\beta^{nor}}{{\mbox {Tr}}
\tilde{\hat{p}}^{q}_q +(1-q) \beta^{nor} \langle \langle \hat{H}
\rangle \rangle_q}\; ,
\end{eqnarray}
where the $\beta^{unnor}$and $\beta^{nor}$ are Lagrange
multipliers associated with  unnormalized and normalized
constraints, respectively.
${ }$

\section{Conclusions}

${ }$ In this work, the path integral formulation for nonextensive
systems based on Tsallis entropy was developed, considering both
the unnormalized and the normalized constraints. For unnormalized
formalism,  we deduced, in a unified way, two different but
equivalent generalizations of the path integral as well as their
related Bloch equations. In the first case, the Bloch equation has
the merit of preserving the form when compared with the usual
Bloch equation, but containing   a temperature-dependent effective
Hamiltonian. This temperature dependence makes the evaluation of
the respective path integral more difficult. The other generalized
Bloch equation has a usual form in the sense that there is no
explicit temperature dependence; but on the other hand, it is a
nonlinear operator equation (it resembles the anomalous correlated
diffusion equation). The corresponding path integral ({\it one and
the same from both generalized Bloch equations}) is obtained from
integral representations of the gamma function. Thus, the usual
path integral ($q=1$) is directly used to perform its
generalization ($q\neq 1$). This fact provides a natural way to
generalize the main properties of the usual path integral,
covering from the usual quantum mechanics to quantum field theory.
Consequently,  the second path integral formulation, based on
integral representations, is more interesting than the first one.
In addition,  we analyzed the path integral formulation using the
normalized constraint in the same way as the unnormalized one,
where the free particle case was analyzed and our result are in
agreement with the classical context \cite{Abe} for one particle
in one-dimensional space. Furthermore,  the free scalar field was
considered.

In particular, this work can be useful in the future investigations
of macroscopic properties  of one-electron states systems with a
multifractal internal structure \cite{quimica}. This idea must be
explored in another opportunity. Let us conclude by saying that we
expect the above generalizations to be useful in the discussion of
generic nonextensive systems, hopefully in a similar way Feynman's
path integral is very useful in many branches of physics.
Moreover, the present generalization appears as a kind of
extension of the scale-invariant diffusion in space-time (see
figure of \cite{feynman1} p. 177), in the sense that it could be
anomalous ($q \ne 1$), instead of the usual Brownian one ($q=1$).
\section*{Acknowledgments}
We gratefully acknowledge useful remarks by L. Borland and
L. R. Evangelista. We also thank
partial financial support by CNPq and PRONEX (Brazilian Agencies).

\appendix \section{}
\label{appendix} Since we have not found  the identity
\begin{equation} \ln (1+x) = \lim_{N\rightarrow \infty}
\sum_{n=1}^N \left(\frac{x}{N}\right) \frac{1}{1+\frac{n x}{N}}
\hskip 1cm (x>-1)
 \label{a1} \end{equation}
 in anywhere else, we
establish it in the following. In order to retain the convergence
in all the steps of the subsequent calculation, we employ a new
variable $y$, defined by the relation  $x=m-1+y$, where $m\geq
1/2$, and $\mid y\mid <m$. Thus, the identity (\ref{a1}) can be
written as
\begin{equation}
\label{a2} \ln (m+y) = \lim_{N\rightarrow \infty}
\sum_{n=1}^N \left(\frac{m-1+y}{N}\right)
\frac{1}{1+\frac{n(m-1+y)}{N}}\; .
\end{equation}
After a simple algebra, we obtain
\begin{equation}\label{a3}  \lim_{N\rightarrow \infty}
\sum_{n=1}^N \left(\frac{m-1+y}{m N}\right)
\sum_{k=0}^\infty (-1)^k a^k \;,
\end{equation} for r.h.s. of (\ref{a2}),
where $a= \frac{y}{m}\frac{n}{N} + \frac{m-1}{m}\frac{(n-N)}{N} <
1$. Furthermore, by using a binomial expansion for $a$, in the
above expression we have
\begin{equation}\label{a4}
\lim_{N\rightarrow \infty} \sum_{k=0}^\infty \sum_{s=0}^k (-1)^k
\left(\begin{array}{c} k \\ s \end{array}\right)
\left(\frac{m-1+y}{m N}\right) \left[ \frac{y^{k-s}(m-1)^s}{(m
N)^k}\right] \sum_{n=1}^N  n^{k-s} (n-N)^s\; .
\end{equation}
To evaluate the sum in $n$, it is convenient to employ a
further binomial expansion,
\begin{equation}
(n-N)^s=\sum_{u=0}^s \left(\begin{array}{c} s \\ u \end{array}\right)
n^{u} (-N)^{s-u}.
\end{equation}
After this, we can sum over $n$, retaining only the dominating
contribution in $N$, as the other contributions vanish in the
limit $N\rightarrow\infty$. In this way, r.h.s. of (\ref{a2}) is
reduced to
\begin{equation}\label{a5}
\sum_{k=0}^\infty   \frac{1}{k+1} \left( \frac{m-1+y}{m^{k+1}}   \right)
\sum_{s=0}^k (-1)^s (m-1)^s y^{x-s} \;. \end{equation}
Performing the sum over $s=0$, we  get for this expression
\begin{equation}\label{a6}
\sum_{k=0}^\infty \frac{(-1)^k}{k+1}  \left(\frac{y}{m} \right)^{k+1} +
\sum_{k=1}^{\infty} \frac{1}{k} \left(\frac{m-1}{m}\right)^k \; .
\end{equation}
The resultant  sums are respectively the series representations
for $ \ln \left( 1+\frac{y}{m} \right)$ and $\ln (m)$,
establishing, therefore, the identity (\ref{a2}).


\end{document}